# Stable Deuterium-Tritium burning plasmas with improved confinement in the presence of energetic-ion instabilities.

**Authors**

Jeronimo Garcia[1*], Yevgen Kazakov[2], Rui Coelho[3], Mykola Dreval[4], Elena de la Luna [5], Emilia R. Solano[5], Ziga Stancar [6], Jacobo Varela[7], Matteo Baruzzo[8], Emily Belli[9], Phillip J. Bonofiglo[10], Jeff Candy[9], Costanza F. Maggi[6], Joelle Mailloux[6], Samuele Mazzi[1], Jef Ongena[2], Michal Poradzinski[6], Juan R. Ruiz[11], Sergei Sharapov[6], David Zarzoso[12] and JET contributors[a]

**Affiliations**

[1]CEA, IRFM, F-13108 Saint-Paul-lez-Durance, France.

[2]Laboratory for Plasma Physics, LPP-ERM/KMS, EUROfusion Consortium member, TEC Partner, Brussels, Belgium.

[3]Instituto de Plasmas e Fusao Nuclear, Instituto Superior Técnico, Universidade de Lisboa, Portugal.

[4]National Science Center Kharkiv Institute of Physics and Technology, 1 Akademichna Str., Kharkiv 61108, Ukraine.

[5]Laboratorio Nacional de Fusion, CIEMAT, 28040 Madrid, Spain.

[6]United Kingdom Atomic Energy Authority, Culham Science Centre, Abingdon, Oxon OX14 3DB, United Kingdom of Great Britain and Northern Ireland.

[7]Universidad Carlos III de Madrid, 28911 Leganes, Madrid, Spain.

[8]Dip.to Fusione e Tecnologie per la Sicurezza Nucleare, ENEA C. R. Frascati, via E. Fermi 45, 00044 Frascati (Roma), Italy.

[9]General Atomics, PO Box 85608, San Diego, CA 92186-5608, United States of America.

[10]Princeton Plasma Physics Laboratory, Princeton, New Jersey 08534, USA.

[11]Rudolf Peierls Centre for Theoretical Physics, University of Oxford, Oxford OX1 3NP, United Kingdom.

[12]Aix-Marseille Université, CNRS, Centrale Marseille, M2P2, UMR 7340 Marseille, France.

[a] See the author list of "Overview of T and D-T results in JET with ITER-like wall" by CF Maggi et al. to be published in Nuclear Fusion Special Issue: Overview and Summary Papers from the 29th Fusion Energy Conference (London, UK, 16-21 October 2023)

**Abstract**

Providing stable and clean energy sources is a necessity for the increasing demands of humanity. Energy produced by fusion reactions, in particular in tokamaks, is a promising path towards that goal. However, there is little experience with plasmas under conditions close to those expected in future fusion reactors, because it requires the fusion of Deuterium (D) and Tritium (T), while most of the experiments are currently performed in pure D. After more than 20 years, the Joint European Torus (JET) has carried out new D-T experiments with the aim of exploring the unique characteristics of burning D-T plasmas, such as the presence of highly energetic ions. A new stable, high confinement and impurity-free D-T regime, with strong reduction of energy losses with respect to D, has been found. Multiscale physics mechanisms critically determine the thermal confinement and the fusion power yield. These crucial achievements importantly contribute to the establishment of fusion energy generation as an alternative to fossil fuels.

Corresponding author email: jeronimo.garcia@cea.fr

## Introduction

Modern societies are eager to increase their energy resources, which currently are largely provided by fossil fuels. In this context, plasmas, i.e. the fourth state of matter, which are characterized by the presence of free charged particles, can provide a carbon-free energy source through fusion reactions of light atom nuclei. For that purpose, plasmas must be well confined and reach high pressure in order to overcome the electrostatic repulsion of particles. Magnetic containment is one of the most promising routes towards this goal. The tokamak concept, in which plasmas are confined both by a high toroidal electric current, $I_p$, and a toroidal magnetic field, $B_T$, is particularly promising. Tokamaks have made great progress from their initial proposal [1] [2]. Plasmas with good thermal confinement have been achieved in high-performance tokamak discharges when turbulence-driven energy losses (one of the main physics mechanisms by which plasmas lose their thermal confinement) are strongly reduced. This is the case for the H-modes [3] or internal transport barrier (ITB) regimes [4], in which steep pressure gradients are formed, leading to high plasma pressure. Recently, the EAST [5] and KSTAR [6] tokamaks have reported significant advances. Long-sustained high-performance plasmas were obtained in conditions of low turbulent heat transport with simultaneous avoidance of dangerous magnetohydrodynamic (MHD) bursts, called edge localized modes (ELMs) [7], which are characteristic of H-modes. Such bursts can lead to rapid expulsion of edge plasma and hence to high levels of heat and particle flux to the tokamak wall.

On the one hand, these advances have clarified the route towards a potential fusion energy commercial reactor. On the other hand, the results obtained are not enough to provide a clear insight into how future energy-producing reactor plasmas are expected to behave. The reason is that in significant contrast to plasmas produced nowadays, formed almost exclusively by pure Deuterium (D), energy-producing plasmas will use Deuterium and Tritium (T) to produce 14.1 MeV neutrons and 3.5 MeV $^4$He nuclei (alpha particles). The presence of T has been identified as a potential source of significant changes with respect to pure D plasmas, particularly in terms of turbulence or MHD characteristics [8] [9]. Furthermore, the presence of a small and yet very energetic population of alpha particles, which will provide the main self-heating mechanism through collisions to the thermal plasma, leads to conditions that are mostly unexplored and have been identified as potentially detrimental. The highly energetic fusion-born alpha particles mainly transfer energy to electrons (rather than to thermal ions) by collisions, thus providing strong electron heating. Conditions in which the electron temperature, $T_e$, is significantly higher than the ion temperature, $T_i$, have been identified as leading to unfavorable destabilization of turbulence and clamping of the ion temperatures at very low values, as shown in plasmas with pure electron heating by means of electromagnetic waves [10] [5]. Moreover, alpha particles do not provide significant torque and therefore lead to low toroidal rotation, reducing rotation-driven suppression of turbulence [11][12]. Finally, they can resonantly destabilize Alfvén waves that can produce stochastic transport of alpha particles and hence reduction of fusion energy generation [13][14][15][16].

These novel characteristics are in contrast to present-day experiments which are typically heated by neutral beam injection (NBI) at energies ∼ 100 keV. NBI dominantly heats ions, not electrons, and provides significant torque. Unlike fusion-generated alpha particles, they produce a sizeable fast-ion density fraction with $T_i/T_e > 1$. In such conditions, it is well known that high confinement can be obtained [6].

The Joint European Torus (JET) tokamak [17], which is the only tokamak in the world capable of operating with T, has undergone a new experimental campaign in D-T, DTE2 [18], with the aim of providing solid evidence on the characteristics of D-T plasmas. A world record of fusion energy was produced in JET at the end of 2021 [19] by developing H-mode plasmas with ELMs. However, such an impressive result was obtained under conditions of high NBI power. In this paper, we describe DTE2 plasmas that, rather than maximizing the production of fusion energy, aim to capture other important characteristics expected in future fusion reactors, such as the simultaneous development of conditions with dominant total heating fraction transferred to the electrons, low input torque, and the triggering of Alfvén wave instabilities. These characteristics were not studied in the DTE2 fusion energy record, nor in the first D-T campaign in JET (DTE1)[9] or TFTR[8] in the 90s, since in both cases the maximization of the fusion power by using high NBI power was mostly explored and in such conditions plasmas with $T_i/T_e > 1$ and high ion heating and rotation are obtained. Therefore, the study presented in this paper represents the first time that this path has been pursued in D-T plasmas.

We show that very good properties are achieved in terms of energy confinement and stability. The type of confinement expected in baseline reactor plasmas is obtained because of the better energy confinement in D-T with respect to the same conditions in D. In particular, for the electrons, energy losses by transport are low, which allows temperatures of about 110 million K. For the ions, the core heat transport is significantly reduced in D-T compared to D when instabilities generated by the energetic ions are observed. Under these conditions, up to 1 MW of fusion power is generated with only 2.4 MW of external ion heating. This type of plasma provides an integrated solution for future tokamak reactors, since, despite developing an H-mode, deleterious ELMs are avoided. The results indicate improved energy confinement in D-T burning plasma conditions.

## D-T plasmas development and comparison to D

Several plasma configurations and D-T concentrations were explored in JET to cover a wide range of possible configurations in future tokamak reactors. To minimize the external torque and hence the toroidal rotation, the plasmas presented in this study were mostly heated with electromagnetic waves resonating at the ion cyclotron frequency (ICRF), which ensures that a low external torque is applied. ICRF heating can accelerate ions to MeV energies and in these plasmas 1% of H was used as a minority wave resonator. Additional lower heating power levels of NBI were used with the aim of reaching high temperatures. The discharge #99896, shown in Fig.1A represents the type of plasma performed. The chosen D-T concentration was $\sim 50\%D - 50\%T$ as it is expected to deliver the maximum fusion power in future tokamak fusion reactors.

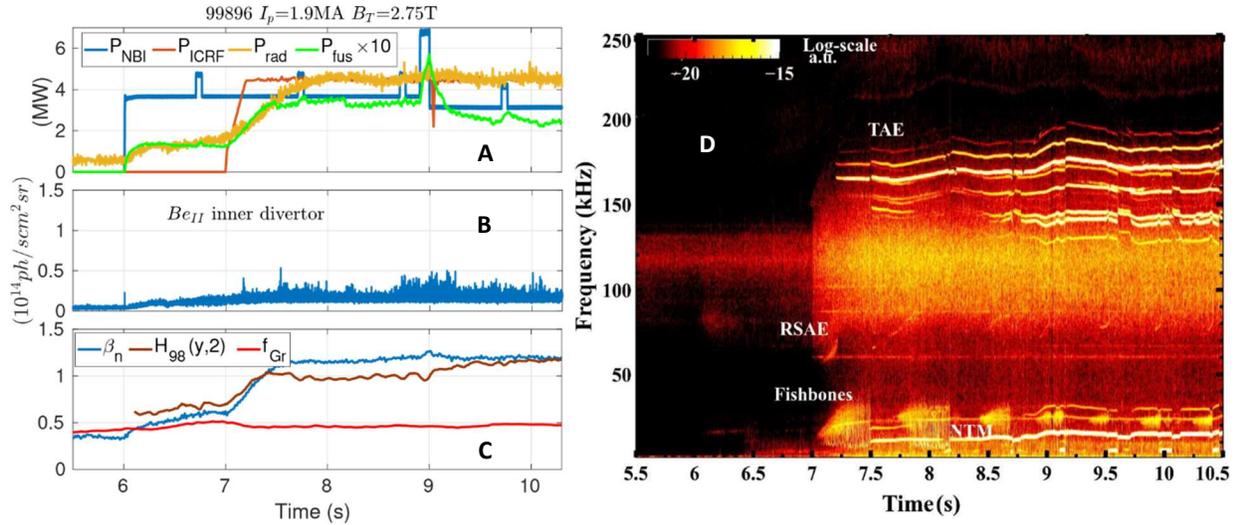

**Fig. 1. Main characteristics of the D-T discharge #99896.** (A) Time evolution of discharge #99896, with toroidal current $I_p$ = 1.9 MA, magnetic field $B_T$ = 2.75 T, and $q_{95}$ = 4.5, heated mainly with ICRF power, $P_{ICRF} = 4.5MW$. The NBI power, $P_{NBI} \sim 3.5MW$, was also injected with deuterium beams, before 9 s, and tritium beams, after 9 s. The radiated power, $P_{rad}$ represents 60% of the total input power. The fusion power obtained reaches a maximum of $P_{fus} \sim 0.5MW$. (B) Time evolution of edge fluctuations as obtained from the $Be_{II}$ line emission from the inner divertor. (C) Time evolution of $\beta_N$, defined as $\beta_N = \beta aB_T/I_p[\%]$ with $\beta$ the ratio between magnetic and thermal

pressure and $a$ the plasma minor radius, $H_{98}(y,2)$ and $f_{Gr} = \bar{n}_e/n_{Gr}$ the Greenwald fraction with $\bar{n}_e$ the average density and $n_{Gr}$ the Greenwald density defined as $n_{Gr} = I_p/\pi a^2$. (D) Time evolution of magnetic perturbations detected by the Mirnov coils.

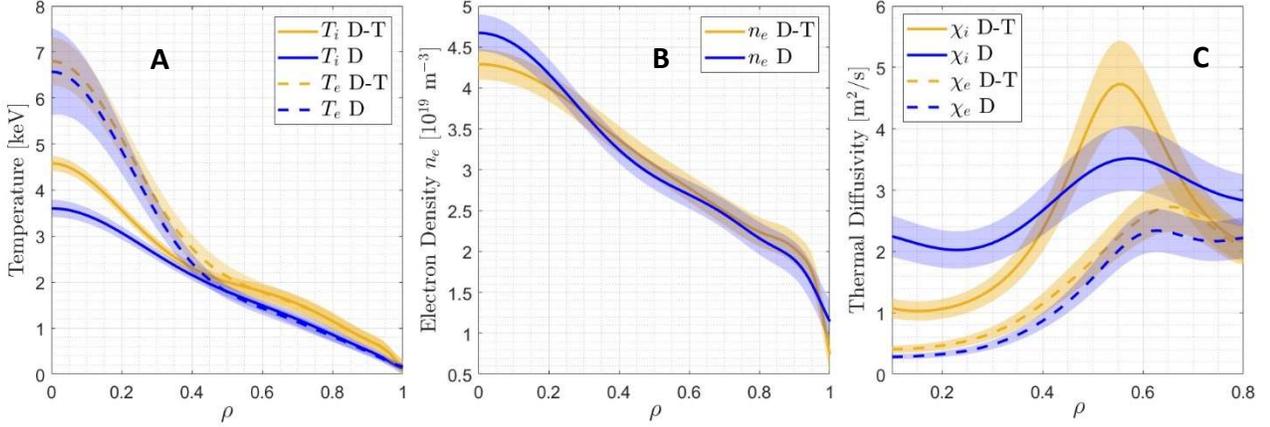

**Fig. 2. Comparison between the D-T discharge #99896 and the D counterpart #100871**. (A) Comparison between $T_i$ and $T_e$ for the D-T discharge #99896 and the D counterpart #100871. $T_i$ is measured by the charge-exchange technique on impurity ions. $T_e$ is obtained by means of LIDAR and high-resolution Thomsom scattering (HRTS). An average over 8.5 s-8.7 s is performed. $\rho$ is defined as the square root of the normalized toroidal magnetic flux. (B) Comparison between electron density, $n_e$, for the D-T discharge #99896 and the D counterpart #100871. $n_e$ is measured with HRTS. (C) Comparison between $\chi_i$ and $\chi_e$ obtained by power balance analysis for the D-T discharge #99896 and the D counterpart #100871.

During the phase with ICRF and NBI heating, the thermal confinement time of this plasma, $\tau$, calculated excluding the contribution of energetic ions, is the reference in the ITER baseline conditions as $H_{98}(y,2) = \tau/\tau_{IPB98} \geq 1$ (Fig.1C) with $\tau_{IPB98}$ the energy confinement time predicted by the IPB98(y,2) scaling (20). This result is obtained under low rotation conditions, since the Mach number at $\rho = 0.5$, $M = v_{tor}/c_s \sim 0.15$, with $v_{tor}$ the plasma rotation, $c_s = \sqrt{T_e/m_i}$ the sound speed and $m_i$ the Deuterium ion mass, is lower than that expected in ITER D-T plasmas (21).

As shown in Fig.1B these results are obtained in the presence of small edge fluctuations rather than ELMs with $\beta_N = 1.2$ and $f_{Gr} \sim 0.45$ (Fig.1C). Analysis using the TRANSP suite of modeling codes (22) indicates that 56% of the total auxiliary heating is transferred to electrons. The high electron heating is confirmed by the observation that $T_e > T_i$, notably in the region $\rho < 0.4$ with $T_e/T_i \sim 1.4$ on the magnetic axis. On the contrary, during the NBI only phase, $H_{98}(y,2) = 0.7$, indicating a lower confinement compared to the phase with injected ICRF heating. In terms of fusion power, $P_{fus} \sim 0.5$ MW was obtained. Finally, the radiated power reaches 60% of the total power injected and is located at the plasma separatrix close to the X-point, which indicates that the accumulation of core impurities, usually a concern in high confinement plasmas in a metallic wall environment such as in JET (23), is avoided.

The good thermal confinement of discharge #99896 is obtained in the presence of generally-believed deleterious electromagnetic perturbations over a wide range of frequencies when ICRF heating is added, as shown in Fig.1D. We can identify perturbations generated by the interplay between energetic ions and Alfvén waves, such as toroidal Alfvén eigenmodes (TAE) (24, 25) and reversed-shear Alfvén eigenmodes (RSAE) (26). Fishbone instabilities (27)(28) related to the interaction between energetic ions and MHD are also present. All of this activity corroborates the presence of highly energetic ions in the plasma. MHD activity and in particular neoclassical tearing modes (NTM) (29)(30) are also found.

An equivalent discharge, #100871, in terms of input power, radiated power, and density, with pure D was also carried out to compare to the characteristics of D-T discharges. The match was successful, including a similar pattern of magnetic instabilities. The comparison between the temperature and electron density profiles between D-T and D is shown in Fig.2(A and B). Although the electron density is nearly identical in D-T and D, both $T_e$ and $T_i$ are higher in D-T for the entire plasma radius. In particular, in the plasma core, an increase in $T_i$ slope is obtained. Since the heating deposition profiles are very similar in D-T and D in the plasma inner core, as verified with TRANSP, such a change in $T_i$ could have its origin in differences in thermal transport. Indeed, a significant change is found, as demonstrated by calculating the power balance heat diffusivity for ions, $\chi_i$, and electrons, $\chi_e$, (see Fig. 2C). In D-T, $\chi_i$ shows a drop in the plasma core starting from $\rho = 0.4$ and reaching $\chi_{i,D-T}/\chi_{i,D} \sim 0.5$ at $\rho = 0.2$. The electron heat diffusivity, $\chi_e$, remains very low for D-T and D, which leads to very peaked $T_e$, similar to ITB plasmas observed in other high electron heated plasmas such as the super I-mode developed in EAST (5). Importantly, $\chi_{i,D-T}/\chi_{e,D-T} \sim 2$ while it is doubled for pure D indicating deteriorated ion thermal confinement.

D-T density-ratio scans were performed to evaluate the impact of T on plasma characteristics. As an example, the discharge #99817 was performed in conditions similar to #99896 but with 85% T fraction and $B_T = 3.7$ T. The total heating to electrons under these conditions was 70% as obtained from TRANSP, with $P_{fus} \sim 1$ MW and $T_e$ reaching 110 million K and $T_i$ 60 million K with only 8 MW of input power. Similar to discharge #99896, a broad range of energetic-ion instabilities was obtained. Compared to the case with $\sim 50\%D - 50\%T$, $\chi_{i,D-T}/\chi_{e,D-T}$ is reduced to $\sim 1$, which means that the presence of T in the plasma is a key player in reducing ion thermal energy losses due to heat transport.

## Core instabilities analyses

Beyond large-scale MHD and Alfvén-driven fluctuations, small-scale microturbulence driven by the Ion Temperature Gradient (ITG) instability [31] is one of the major threats to plasma thermal confinement. In JET plasmas, ITG modes, destabilized at ion-gyroradius scales, $\sim 1$ cm, are usually responsible for enhancing core radial energy transport even in strongly electron-heated plasmas [32]. Instead, energy transport driven by instabilities at electron gyroradius scales, $\sim 0.1$ mm, i.e. by electron temperature gradient modes (ETG), is found to not contribute significantly to overall outward transport in the plasma core [33]. In this section, an analysis of the fluctuations found in plasma #99896 is given, and their role in producing good thermal confinement is clarified.

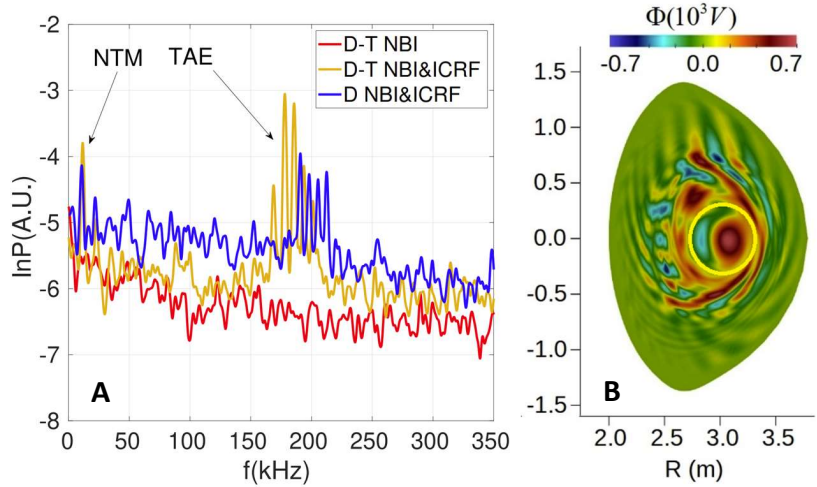

**Figure 3: Core plasma fluctuations characteristics**. (A) Density fluctuations as obtained from reflectometry at major radius $R \sim 3.36$ m, $\rho \sim 0.35$ and $t = 8.4$ s, for the D-T discharge #99896 with only NBI heating or with full NBI and ICRF heating and comparison to the pure D discharge #100871. (B) Electrostatic potential, $\Phi$, fluctuations obtained for the D-T discharge #99896 by the global code FAR3D when considering two species of energetic ions, H and D, accelerated by the ICRF power. The yellow circle represents the $q = 1$ surface. Inside $q = 1$, a $n = 1$ perturbation is obtained which is identified as a fishbone instability. Outside $q = 1$, TAEs are obtained.

A Fourier analysis of the measured magnetic fluctuations is shown in the supplementary information, Fig. S1(A and B). A full range of activity is found at both low and high frequencies. In the frequency range 1 kHz $< f <$ 40 kHz, modes with toroidal mode number $n = 0 - 7$ are detected, while for 120 kHz $< f <$ 200 kHz the modes detected cover $n = -5$ to $n = 6$. However, nonlinear mode-mode interactions are at the origin of some of these fluctuations. This is demonstrated by performing a mode-mode bi-coherence analysis [34]. In particular, the nonlinear interplay between high-frequency TAE and low-frequency NTM is at the origin of the high frequency perturbations with toroidal mode number $n = 1$ and $n = 2$ among other interactions (this is shown in the supplementary information, Fig.2 S2). A necessary condition for such an interaction to occur is that the radial locations of TAE and NTM are close to each other. This requirement is further verified by the analysis of the location of the different perturbations in real space using several techniques, as clarified in the methods section. RSAE and fishbones are located inside $q = 1$, at $\rho \sim 0.25$, with $q$ the safety factor defined as $q \equiv d\Psi_t/d\Psi_p$ with $\Psi_t$ the toroidal magnetic flux and $\Psi_p$ the poloidal magnetic flux. RSAE are destabilized very close to the magnetic axis. TAE and NTM are located just outside $q = 1$, at $0.25 < \rho < 0.45$.

Further evidence of radial and temporal overlap of MHD and TAE beyond $q = 1$, located at $R \sim 3.25$ m, is obtained from density fluctuations using a reflectometry diagnostic [35] at $R \sim 3.36$ m. It is shown in Fig.3A that high frequency fluctuations for both D-T and D are detected in the TAE and NTM frequency ranges. Furthermore, except for the frequencies corresponding to TAE and NTM, density fluctuations are lower for D-T than for D, notably in the typical range of ITG fluctuations $f < 150$ kHz. This supports the improvement in confinement for the thermal ions shown in Section 2. Density fluctuations do not increase significantly when 4.5 MW of ICRF power is added on top of NBI power, except at the TAE and NTM frequencies. This is important because, in general, turbulence increases with increasing input power leading to the so-called thermal confinement degradation with input power [12].

The origin of the destabilization of magnetic fluctuations by energetic ions has been analyzed with the global gyrofluid code FAR3D [36]. Two energetic ion species, H and D, were considered because they are both ICRF accelerated by means of the first and second harmonic absorption. Their characteristics are obtained from TRANSP. The frequency and location of the fishbone and TAE perturbations obtained from linear simulations with FAR3D agree with experimental data as shown in the supplementary information Fig. S3 and S4, which clarifies that the perturbations are destabilized by ICRF-accelerated ions. Nonlinear simulations including both energetic ion species are shown in Fig.3B. The radial extension of the electrostatic potential perturbation, 2.4 m $< R <$ 3.5 m, agrees well with the experimental location and clearly shows that the energetic-ion-induced perturbations extend up to mid-radius, coinciding with the radial extension of the decreased thermal energy transport losses in D-T compared to D.

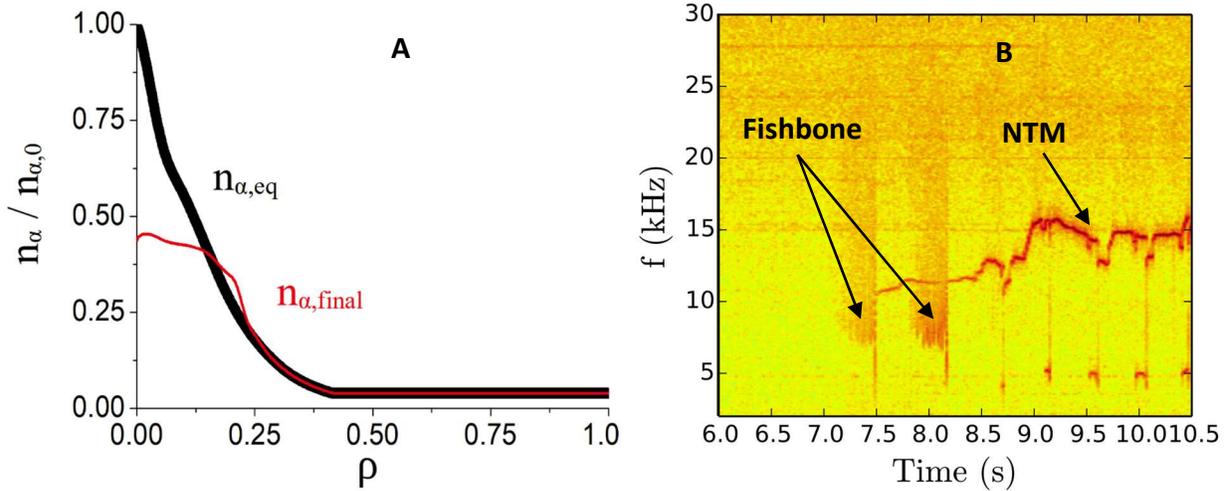

**Figure 4: Alpha particle transport and losses in D-T**. (A) Alpha particle density profile as calculated with TRANSP assuming no alpha particle transport ($n_{\alpha,eq}$) and comparison to the profile obtained from the FAR3D code after the full development of the fishbone instability ($n_{\alpha,final}$). (B) Alpha particle loss frequency spectrum obtained by using the fast ion loss detector (FILD) with channels that are receptive to 3.5 MeV alpha particles.

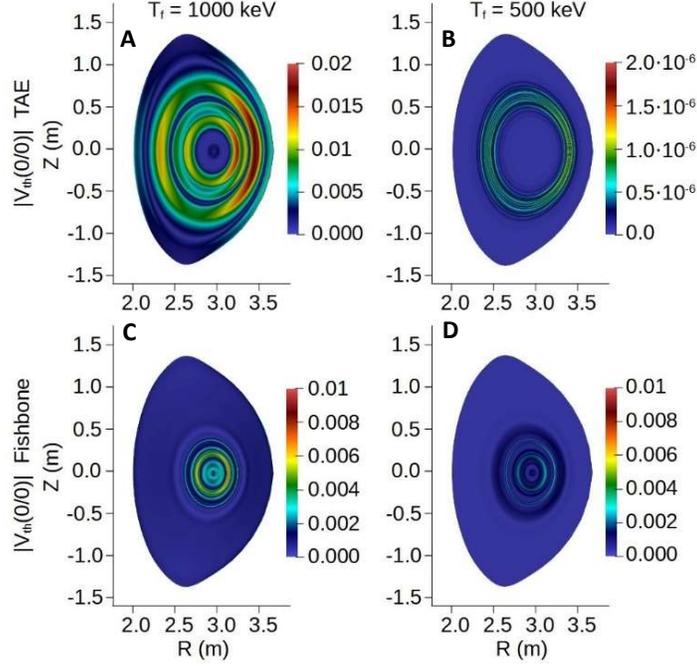

**Figure 5: Zonal flow generation by energetic particle instabilities.** 2D pattern of $n = 0$, $m = 0$ structures of zonal poloidal flows, $V_{th}(0,0)$, for the TAE and fishbone instabilities. $V_{th}(0,0)$ is defined as $V_{th}(0,0) = E_r(0,0) \times B\varphi$ with $E_r(0,0)$ the $n,m = 0,0$ component of the perturbed radial electric field and $B\varphi$ the toroidal magnetic field. The dependence of the zonal flow intensity on the perturbation strength is studied by scanning the energetic ion equivalent temperature ($T_f$) using two values, $T_f = 1$ MeV and $T_f = 500$ keV in the FAR3D code. Zonal flow generation increases with increasing perturbation intensity for both TAEs (A and B) and fishbones (C and D). The radial extension of zonal flow activity coincides with the extension of the two perturbations.

In addition to the nonlinear interplay between TAE and NTM, other complex and multi-scale interactions are of paramount interest and critically determine the performance. It is found with FAR3D that alpha particles do not destabilize any perturbation due to their low density, however, nonlinear interplay with fishbones can induce radial transport and losses of alpha particles, partially depleting the plasma axis of such particles as shown in Fig.4A. This is experimentally corroborated, as shown in Fig.4B, by means of the fast ion loss detector (FILD) [37]. In the initial phase of the discharge, when the activity of the fishbone is especially strong, alpha particle losses are detected, as can be seen at $t \sim 8$ s. In the later phase of the discharge, fishbone activity is reduced in intensity and no further losses are detected with origin in the fishbone perturbation, but rather in the NTM at $f \sim 15$ kHz. Unlike fishbones, no alpha losses with origin in TAE are detected both in the FAR3D simulations and in the experiment. Therefore, it becomes clear that although magnetic perturbations do not prevent reaching high confinement, they can lead to loss of fusion power. These results show the critical interaction between magnetic perturbations and alpha particles, and we conclude that it is essential to control such interactions in order to produce high fusion power in future tokamak reactors.

Nevertheless, magnetic perturbations induced by energetic ions can also lead to beneficial effects that may have a positive impact on reducing energy losses by heat transport. This is the case of the interplay with the so-called zonal flows, i.e. thermal plasma flows with $f \sim 0$ and poloidal and toroidal perturbation mode numbers, $n,m = 0,0$. Zonal flows were theoretically predicted [38] and also obtained in dedicated numerical simulations in the presence of energetic ions [36][39][40]. They are known to reduce transport driven by ITG turbulence [41], in particular in the presence of energetic MeV ions as shown in D-$^3$He plasmas [32][42].

The generation of zonal flows is studied for discharge #99896 by analyzing the energy transfer between fishbone and TAE perturbations and the thermal plasma. The 2D pattern of n = 0 structures for zonal poloidal flows, as obtained from FAR3D nonlinear simulations, are shown in Fig.5(A to D) considering two different energies for the energetic ions. Clearly, fishbones and TAE drive zonal flows, with higher intensity with increased energetic ion energy, and hence stronger instability drive.

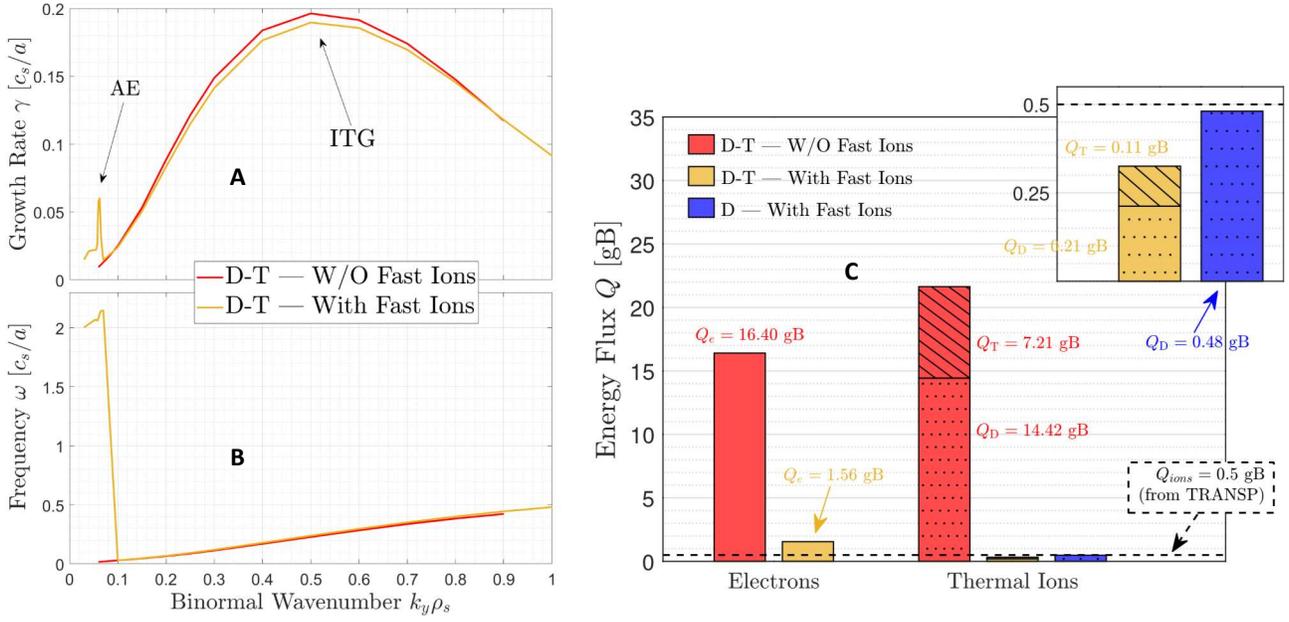

**Figure 6: Gyrokinetic analysis of core turbulence reduction in D-T**. (A) Growth rate, $\gamma$, and (B) frequency, $\omega$, spectrum obtained from linear simulations with the CGYRO code. $k_y$ is the binormal wavenumber normalized to the proton sound gyroradius $\rho_s$. (C) Energy flux obtained from gyrokinetic simulations performed with the CGYRO code for the discharge #99896 at $\rho$ = 0.31. The simulations are performed including and excluding the energetic ion component. Values of the ion thermal energy flux deduced from power balance in TRANSP (black horizontal dashed line) are only obtained when the energetic ion component is included in the simulations as a separate species. The total thermal ion energy flux obtained in D-T including energetic ions is compared to the one obtained assuming that all the thermal ions are D while keeping the rest of the parameters fixed. The energy flux in pure D is significantly higher than in D-T (a zoom of those bars is displayed in the inset on the top right).

The role of T and energetic ions on the good plasma confinement of discharge #99896 is further analyzed by performing simulations with CGYRO [43]. CGYRO solves the gyrokinetic-Maxwell equations [44] to obtain the electrostatic and electromagnetic fluctuations and corresponding turbulent energy transport. Local simulations are performed at $\rho$ = 0.31 by including and excluding energetic ions as separate species in addition to the electrons, D and T species. Due to the low energetic ion density compared to the electron density, $\sim$ 3%, the growth rates obtained in linear simulations are nearly unaffected in the ITG scales by the presence of energetic ions as shown in Fig.6(A and B). However, low $k_y$ modes with TAE frequency are destabilized. These results show that the type of plasma found in JET is different to other plasmas dominated by energetic ions effects, such as the FIRE mode [6], which is characterized by strong turbulence reduction with energetic ion dilution and linear effects [45].

Regarding non-linear effects, as shown in Fig. 6(C), the energy flux, $Q$, obtained when including energetic ions is more than ten times lower and is close to the values calculated from power balance analysis from TRANSP. Importantly, such a strong reduction in thermal energy flux is accompanied by a high increase in zonal flow shearing activity, $\omega_{E\times B}$, which is five times higher when energetic ions are included and thus confirming the results obtained with FAR3D. Furthermore, there is a clear asymmetry between the transport obtained in D and in T, with the T transport systematically lower than that for D, $\chi_{i,T} \sim 0.8 \chi_{i,D}$. Such a difference has an important consequence on the total flux in D-T compared to that of pure D. This is numerically analyzed by performing an alternative simulation in which the T ions are artificially considered as D, thus performing a pure D simulation. The turbulent energy flux ratio $Q_{D-T}/Q_D$ = 0.67 is similar to the power balance one obtained for the discharges #100871 in D and #99896 in D-T at the same radial location, $Q_{99896}/Q_{100871}$ = 0.71. This result confirms expectations from purely numerical simulations performed for D-T plasmas [46][47][48]. From the numerical point of view, global effects from profile shearing were investigated in CGYRO and found them to negligibly affect the thermal fluxes at the radial location studied.

In summary, the analyses of core plasma fluctuations indicate an optimum route towards the generation of fusion power in D-T tokamak plasmas whereby energetic ion instabilities produced by alpha particles remain in conditions of negligible or weak alpha particle transport, while they can induce thermal energy transport reduction by means of zonal flows. Importantly, this is obtained in conditions of some energetic ions characteristics relevant to ITER burning plasmas, e.g. the energetic ion density is similar to the one expected in ITER, $\sim 1\%$ [47], avoiding energetic ion dilution as usually happens in strong NBI heated plasmas.

## Pedestal formation in D-T

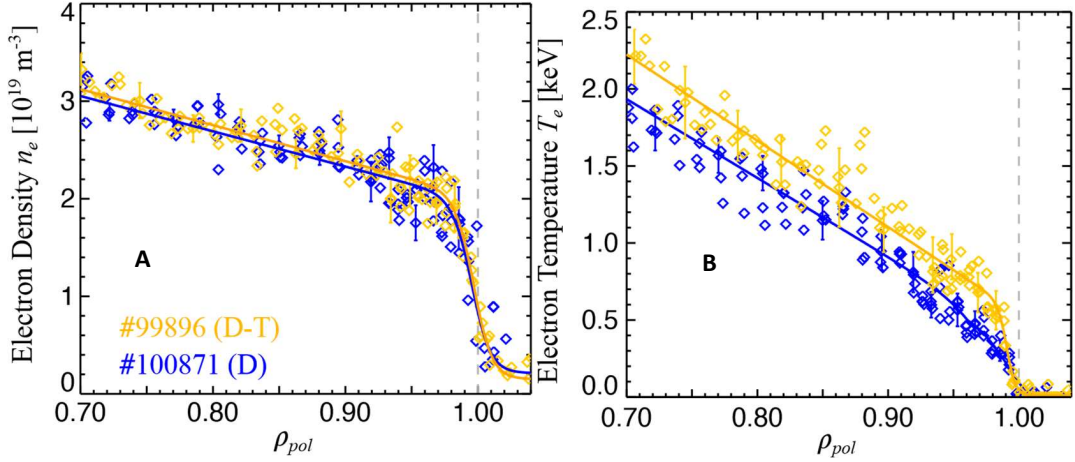

**Figure 7: Pedestal formation in D-T and comparison to D.** (A) Comparison between edge $n_e$ for the D-T discharge #99896 and D discharge #100871. $\rho_{pol}$ is defined as the normalized poloidal flux. (B) Comparison between edge $T_e$ for the D-T discharge #99896 and D discharge #100871. The vertical dashed line represents the location of the plasma separatrix. The profiles are obtained from HRTS averaged in the time window 8.5 s-8.9 s for #99896 and 8.4 s-8.8 s for #100871.

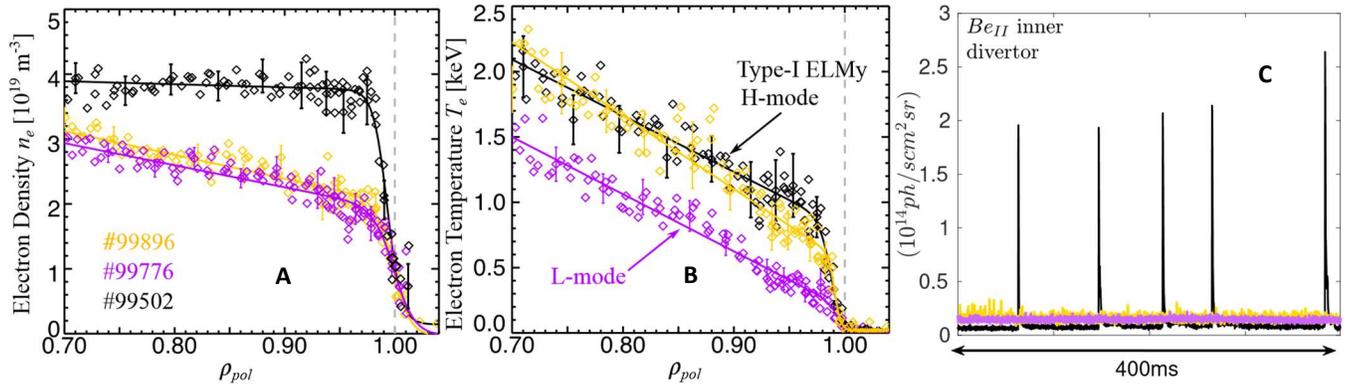

**Figure 8: Pedestal characteristics in D-T.** Comparison between discharge #99896 and discharge #99502, with $P_{NBI}$ = 12.5 MW, in H-mode with type-I ELMs, and #99776, in L-mode, with $P_{NBI}$ = 5.4 MW and $P_{ICRF}$ = 3.3 MW both obtained at $I_p$=2.5 MA, $B_T$=3.7 T and $q_{95}$=4.5. (A) Edge $n_e$. (B) Edge $T_e$. The profiles are obtained from HRTS averaged in the time window 8.5 s-8.9 s for #99896, 7.3 s-7.6 s for #99502 and 8.6 s-8.9 s for #99776. (C) Comparison of the BeII line emission from the inner divertor for the same discharges. The vertical dashed line represents the location of the plasma separatrix.

As depicted in Fig.7 (A and B), the D-T discharge #99896 shows the formation of an H-mode with a steep edge temperature gradient, i.e. a pedestal, at $\rho_{pol} \sim 0.95$. The temperature of the D counterpart is lower at the same location, while the density is nearly identical. The fact that the pedestal pressure is higher with increasing isotope mass has been observed in H-mode plasmas with ELMs (ELMy H-mode). However, in ELMy H-mode regimes with typical type-I ELMs [49][50][51], an increase in edge density rather than temperature is obtained with increasing isotope mass [52].

With the aim of further investigating the origin of the pedestal found in discharge #99896, it is compared to two D-T discharges, one with a clear transition to H-mode with only NBI power and the development of typical type-I ELMy H-mode and another, with NBI and ICRF heating that remains in L-mode, i.e. without temperature pedestal. As shown in Fig.8A, the edge density of the discharges #99896 and the one in L-mode is nearly identical, demonstrating that the density of discharge #99896 remains in L-mode, yet the pedestal formation is evident as the edge temperature nearly reaches that of the ELMy plasma (Fig.8B). Importantly, discharge #99896 has no ELMs as shown in Fig.8C in which the divertor oscillations from BeII divertor emission are compared to those obtained in L-mode and with ELMs. Clearly, the fluctuations are closer to those obtained in L-mode.

The spontaneous generation of plasmas with no ELMs and a pedestal for the temperature is systematic in these kinds of plasmas performed in D-T at different $I_p$ and $B_T$. It is apparent from the available data that, in those plasma conditions, the input power is close to the L to H-mode transition power threshold, but below the power required to fully develop ELMs. This is supported by the fact that during the phase where the temperature pedestal is sustained an $n = 0$ coherent mode can be observed in the magnetic sensors with a frequency of 5 kHz. This mode, known in JET as M-mode[53], is typically detected in JET immediately after the L to H-mode transition. Long phases with an M-mode present are typically observed in the L-H transition at low density in JET, where pedestal dynamics similar to that described here have been identified [54]. This aspect was further studied in the particular case of discharge #99896, for which it was found that an additional 3 MW of input power lead to the formation of an ELMy H-mode regime. Regarding the magnetic configuration in which these results were obtained, it was used the standard $B \times \nabla B$ in JET with the ion $\nabla B$ drift towards the dominant X-point. This configuration is known to be 'favourable' in terms of power requirements for the access to H-mode.

These plasmas show similarities with the I-mode regime [55][56][57], found in D, which is characterized by the formation of a pedestal for the temperature, while the density remains in L-mode. Similar to the I-mode, the no-ELM regime described here has good impurity transport properties, with no impurity accumulation. However, no signs of the so-called weakly coherent mode (WCM), typically found in I-mode plasmas, have been detected.

## An scenario towards D-T burning plasmas

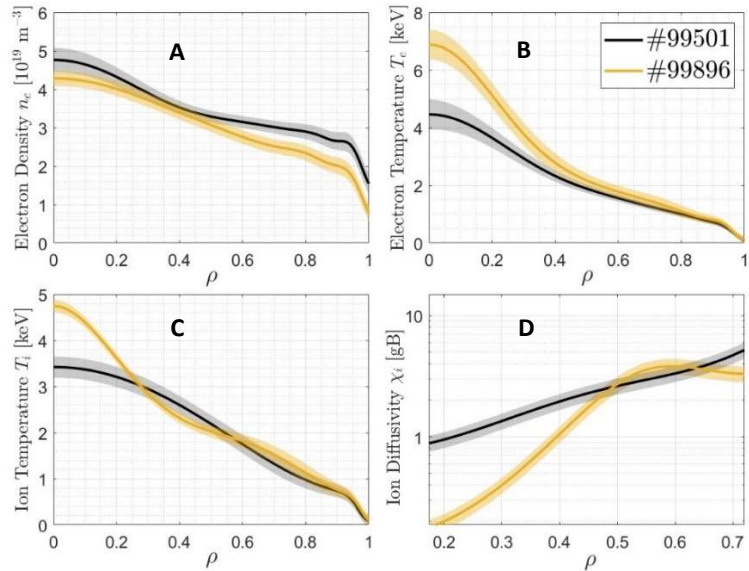

**Figure 9: Scenario performance in D-T**. Comparison between the discharge #99896, obtained at $P_{NBI}$=3.5 MW, $P_{ICRF}$=4.5 MW, $I_p$=1.9 MA, $B_T$=2.8 T and the D-T discharge #99501, obtained at $P_{NBI}$=9.5 MW, $I_p$=2.5 MA, $B_T$=3.7 T. (A) $n_e$. (B) $T_e$. (C) $T_i$. (D) $\chi_i/\chi_{GB}$. $\chi_{GB}$ is the GyroBohm diffusivity defined as $\chi_{GB} = T_e^{3/2} m_p^{1/2}/(e^2 B_T^2 a)$ with $m_p$ the proton mass, $e$ the electron charge and $a$ the plasma minor radius.

Compared to the more typical H modes that develop ELMs and are heated with NBI power, the D-T discharge #99896 provides an attractive alternative with similar thermal energy, but obtained at lower input power, $I_p$ and $B_T$. Such a feature could lead to a more economical and simpler tokamak design. This is shown in Fig.9(A to C) by comparing the discharge #99896 and the D-T H-mode discharge #99501, at higher $P_{NBI}$, $I_p$, and $B_T$ and heated with pure NBI power. Clearly, the density is higher at the edge for the discharge #99501, since a pedestal is formed for both the density and the temperature and yet the total thermal energy content of the two discharges, $W_{th}$, is similar, $W_{th} \sim 2.4$ MJ. The reason is that the lower edge density is compensated by the higher core temperatures and lower core energy transport losses as shown in Fig.9(D) by comparing $\chi_i$, which is nearly ten times lower in the plasma core for the discharge #99896.

## Discussion

The path towards commercial fusion reactors, although better understood in recent decades, still poses physics and technological uncertainties. The size, magnetic configuration, type of confinement or power exhaust techniques expected in future tokamak devices are not fully established. Therefore, it is of fundamental importance to further clarify a safe and clear route toward the generation of efficient energy by means of fusion reactions. This is especially important for D-T plasmas. The presence of T, the generation of a high neutron rate at 14.1 MeV energy, or the presence of alpha particles, are all characteristics of D-T that are not present in ubiquitous pure-D discharges. Studying the impact of such differences is critical in order to properly characterize how D-T fusion reactors might behave. In particular, T can have a strong impact on confinement, impurity generation, and stability, while alpha particles can lead to significant destabilization of magnetic perturbations and provide electron heating.

The JET tokamak has recently performed a new D-T campaign after the first ones were developed at TFTR [8] and JET [9] in the 90's, in order to address some of these points which were not previously studied. A particular emphasis has been put on the exploration of some of the unique features expected in burning D-T plasmas, i.e. simultaneous high electron heating, low rotation, and fully destabilized energetic ions instabilities.

A major result has been obtained suggesting that some reactor-relevant plasma conditions may be very beneficial. A clear reduction of energy losses by turbulence in D-T with respect to D leads to higher confinement plasmas, free of core impurity accumulation, while avoiding dangerous ELMs typical of standard H-modes and thus providing a stable plasma regime. These results confirm how crucial it is to perform experiments in D-T compared to D, as in addition to changes in confinement, burning D-T plasmas can be characterized by a strong nonlinear interplay between different plasma scales in the presence of highly energetic ions. This interplay can induce alpha particle transport and losses but also induce physical mechanisms, such as zonal flows, which can lead to an improved thermal confinement that can compensate for this effect and increase the fusion power generation.

Our findings pave the way for a more economical and simpler design of tokamaks, confirming that nuclear fusion by means of magnetically confined D-T plasmas is a promising source of clean energy. However, further studies, such as compatibility with power exhaust capabilities and exploration at higher density and power, are required to fully qualify these plasmas as a solid route towards tokamak reactors. Furthermore, it is necessary to perform more detailed modeling activities to analyze multiscale effects involving energetic ions and plasma perturbations at different spatial and temporal scales. Specially important is to develop plasmas in which, unlike in the JET results shown in this paper, alpha particle heating is dominant. To this end, the experimental and modeling efforts in D-T including a significant population of alpha particles, as expected in ITER [58] or SPARC[59], are essential.

## Methods

### Experimental Design

The JET tokamak has investigated some of the most important fusion reactor conditions by conducting a new D-T campaign with Be/W wall. To reproduce the simultaneous high electron heating, low torque, and the destabilization of energetic ions related instabilities expected in future tokamak reactors, JET has mostly used

ICRF power rather than NBI. Several experimental conditions were explored to find an optimum plasma state in terms of confinement, energetic ion production, and D-T fusion power yield. One of the limitations of such exploratory work was the ICRF power, and hence the amount of electron heating available, which was limited to ∼ 4.5 MW, whereas the NBI power was used up to 10 MW in pure NBI plasmas. The ICRF frequency used was 55 MHz at $B_T$= 3.7 T and 42 MHz at $B_T$= 2.75 T.

Different scans were performed for $I_p$ and $B_T$, e.g. $I_p$ was explored in the range 1.9 MA < $I_p$ < 2.5 MA at two $B_T$ = 2.75 T, 3.7 T. The D-T concentration ratio was scanned from 40% T to 85% T by using different valves injecting neutral D and T gases.

**Key diagnostics for D-T operation**

The ion temperature profiles in this paper were obtained from charge exchange recombination spectroscopy (CXRS) [60] measurements of impurity ions and electron temperature profiles from combined analysis of LIDAR Thomson scattering [61] and high-resolution Thomson scattering (HRTS) [62] diagnostics. The density profiles were taken from HRTS measurements, with the density normalized to match the line-average density measured by a far-infrared interferometer.

Mirnov coils are used as a standard MHD diagnostic on almost all tokamak devices. The coils are installed within the vacuum vessel close to the plasma boundary and provide a measurement of the time derivative of the magnetic field. Magnetic spectrograms (Fourier decomposition of the Mirnov coil signal) can then be used to identify relevant oscillation frequencies associated with MHD activity. In JET a number of coil arrays with high-frequency response are available, allowing activity in the Alfvén range to be observed.

The time-resolved neutron yield in JET is measured using three fission chambers, containing $^{235}$U and $^{238}$U, located outside the vacuum vessel.

The Alfvén eigenmode active diagnostic (AEAD) [63] is characterized by six toroidally spaced antennas, each with independent power and phasing, whose aim is to actively excite marginally stable TAEs.

Alpha particle losses are detected by the fast ion loss detector (FILD) consisting of a Faraday cup array [37]. The Faraday cup array is composed of multiple cups that span a wide poloidal angle below the outboard midplane at a single toroidal location with a minor radial extent.

ELMs are characterized by the BeII emission signal from the inner divetor region.

The plasma isotopic composition is measured at the divertor comparing the relative amplitude of Balmer $D_\alpha$ and $T_\alpha$ spectral lines. The D and T ratio in the plasma core is assumed to be equal to the edge, as is usually the case in JET in the presence of multi-ion plasmas when turbulence is driven by ITG [64].

The JET X-mode reflectometry diagnostic [35] is composed of four distinct radial correlation reflectometers. All these reflectometers probe the mid-plane JET plasma. Plasma fluctuations can be obtained from the phase fluctuations of the reflectometer signal.

**Magnetic perturbation spatial location and q profile verification**

The q profile for discharge #99896 has been obtained by means of a loop between the EFIT code and the TRANSP code. The EFIT code calculates the magnetic equilibrium with the input data for the energetic ions content from TRANSP simulations. After a few iterations, a converged q profile is obtained. The validation of the q profile obtained from TRANSP and used for modeling with FAR3D was carried out against a series of diagnostics and MHD markers. As markers with the strongest signature in the diagnostics, the destabilised NTMs were identified and their toroidal mode number calculated using a toroidal array of Mirnov coils. The radial location of the associated rational surface q=4/3 was inferred using two methods: the first uses as proxy the location of the phase inversion of the perturbed electron temperature derived from Electron Cyclotron Emission (ECE) at the NTM frequency, the second matched the NTM frequency as derived from the Mirnov coils to the Doppler-shifted plasma rotation i.e. $n \times V_{phi}$, where $V_{phi}$ is the toroidal rotation of the main plasma ions as derived from CXRS diagnostic. The radial location of the q=1/1 surface was inferred from the inversion radius of the ECE temperature profile during sawtooth crashes as well as from the fishbones signatures in the perturbed plasma temperature from ECE, evidencing a typical kink-like pattern inside the q=1 surface. Lastly, the RSAEs and TAEs were located using Soft X-ray cameras, interferometry and reflectometry.

## Experimental profiles fitting

The profile fitting algorithm makes use of a Gaussian process regression (GPR), which is not limited by a selection of specific fit functions and provides a statistically rigorous estimation of the confidence bounds of the fit. For more details, see the book on the topic written by Carl Edward Rasmussen and Christopher K. I. Williams http://gaussianprocess.org/gpml/chapters/

## TRANSP simulations

The pulses shown in this article were analyzed by interpretive simulations performed with the TRANSP modelling suite [65] coupled with external heating modules NUBEAM (NBI) [66] and TORIC (ICRF) [67], and prepared with the OMFIT integrated modelling platform [68]. Interpretive analysis was based on the use of fitted profiles, including electron density and temperatures. The fitting of $T_e$, $n_e$ and $T_i$ were performed on data obtained from HRTS and CXRS. The fitting of experimental profiles consists on applying a global third-order polynomial fit in the range $\rho \lesssim 0.8$ (with the additional constraint $\partial T_i(0)/\partial r = 0$).

## FAR3D description and simulations parameters

The gyrofluid FAR3D code solves the linear and nonlinear reduced resistive MHD equations describing the thermal plasma evolution coupled with the first two moments of the gyro-kinetic equation, the equations of the energetic particle density and parallel velocity moments [69] [70], introducing the wave-particle resonance effects required for Landau damping/growth. The correct model calibration requires performing gyrokinetic simulations to calculate the Landau closure coefficients in the gyrofluid simulations, matching the analytic TAE growth rates of the two-pole approximation of the plasma dispersion function with a Lorentzian energy distribution function for the energetic particles. The lowest order Lorentzian is matched with a Maxwellian distribution by choosing an equivalent average energy. Further details of the model equations can be found in references [71][72].

A set of linear simulations are performed to reproduce the Alfven eigenmode (AE) activity observed in the discharge, identifying the resonance induced by populations of energetic particles (EP) as passing D and trapped H. The analysis is based on a parametric study with respect to the EP beta (EP density in the plasma) and energy, calculating AEs consistent with the frequency range, plasma radial location, modes number and AE family observed in the experiment. Nonlinear simulations including passing D and trapped H populations are performed to analyze the saturation phase of the AE instabilities, particularly the energetic particle transport induced, the generation of zonal structure, and the nonlinear interaction between different EP populations. The simulations are performed using the EP model profiles obtained from TRANSP, the measured thermal plasma profiles, and the equilibrium calculated with VMEC code [73].

## CGYRO description and simulations parameters

**Table 1:** Employed plasma parameters in CGYRO simulations modelling JET pulse #99896 at $\rho = 0.31$ and $t = 8.6$ s.

| $\epsilon$ | $q$ | $\hat{s}$ | $T_i/T_e$ | $a/L_{n_e}$ | $a/L_{T_e}$ | $a/L_{T_i}$ | $n_D/n_e$ | $n_T/n_e$ | $a/L_{n_D}$ |
|---|---|---|---|---|---|---|---|---|---|
| 0.30 | 1.17 | 0.74 | 0.76 | 1.00 | 3.20 | 2.40 | 0.58 | 0.39 | 0.93 |

| $a/L_{n_T}$ | $n_f/n_e$ | $T_f/T_e$ | $\beta_e$ (%) | $\nu_{ee}a/c_s$ | $B_0$ (T) | $T_e$ (keV) | $n_e$ (m$^{-3}$) | $a$ (m) | $R_0$ (m) |
|---|---|---|---|---|---|---|---|---|---|
| 0.92 | 0.03 | 21.0 | 0.44 | 0.39 | 2.75 | 3.68 | $3.67 \cdot 10^{19}$ | 0.91 | 3.00 |

Here, $\epsilon$ represents the inverse aspect ratio, $q$ the safety factor, $\hat{s}$ the magnetic shear, $n$ the species density normalized to the electron density, $a/L_{n,T}$ the normalized logarithmic density and temperature gradient, $\beta_e$ the electron-beta, $\nu_{ee}$ the electron collision rate and $a$ the minor radius. The normalization factors in standard units are also reported, i.e. the on-axis magnetic field strength $B_0$, the local ($\rho = 0.31$) electron temperature $T_e$, density $n_e$ and the major radius $R_0$.

The CGYRO code [43] solves the electromagnetic gyrokinetic-Maxwell equations [44]. Local simulations were carried out at $\rho = 0.31$. Shaped, up-down symmetric flux-surface geometry was used and multi-species collisions were included using the Sugama collision operator [74]. Transverse and compressional electromagnetic fluctuations were retained. Rotation effects were assumed to be small in the core and not included. Kinetic electrons, D and T as separate species, were included in the simulations. Regarding the energetic-ion species, a lumped H-D species with effective mass averaged between H and D was assumed and modeled by fitting the energetic particle distribution to an equivalent high temperature Maxwellian equilibrium distribution.

The simulations used a radial box length of $L_x = 673\rho_s$ and a binormal box of length $L_y = 628\rho_s$. $N_r = 768$ radial modes and $N_\alpha = 64$ complex toroidal modes were retained. Other resolution parameters were: $N_\theta = 24$ (field line resolution), $N_\xi = 24$ (pitch-angle resolution), $N_u = 8$ (energy resolution) with maximum energy $u^2_{max} = 8$. The definitions of the CGYRO numerical resolution parameters can be found in [43]. The energy flux is provided in GB unit defined as $Q_{GB} = n_e T_e c_s \rho^2_*$, with $c_s = \sqrt{T_e/m_p}$, $\rho_* = \rho_s/a$ is the ratio of the proton sound gyroradius, $\rho_s = c_s/\Omega_c$, to the system size, with $\Omega_c = eB_0/m_p$ the ion gyrofrequency. Convergence test have been performed indicating that the CGYRO results are well resolved.

The zonal flow shearing is defined in CGYRO as:

$$\omega_{E\times B} = \sum_{k_x} k_x^2 \rho_s^2 \langle |\widehat{\Phi}(k_y, k_x)|^2 \rangle_t |_{k_y=0}$$

where $k_x$ is the radial wavenumber, $\widehat{\Phi}(k_y, k_x)$ is the fluctuating electrostatic potential and $\langle\rangle$ denotes the temporal average

The equilibrium profile and geometry parameters are given in Table 1.

**Data availability**

The JET experimental data is stored in the PPF (Processed Pulse File) system which is a centralised data storage and retrieval system for data derived from raw measurements within the JET Torus, and from other sources such as simulation programs. These data are fully available for the EUROfusion consortium members and can be accessed by non-members under request to EUROfusion.

Numerical data that support the outcome of this study are available from the corresponding author upon request.

**Code availability**

The research codes cited in the paper require a prior detailed knowledge of the implemented physics models and are under continuous development. The corresponding author can be contacted for any further information.

Fernandes, J. Ferrand, D.R. Ferreira, J. Ferreira, G. Ferr`o, J. Fessey, O. Ficker, A.R. Field, A. Figueiredo, J. Figueiredo, A. Fil, N. Fil, P. Finburg, D. Fiorucci, U. Fischer, G. Fishpool, L. Fittill,

M. Fitzgerald, D. Flammini, J. Flanagan, K. Flinders, S. Foley, N. Fonnesu, M. Fontana, J.M. Fontdecaba,

S. Forbes, A. Formisano, T. Fornal, L. Fortuna, E. Fortuna-Zalesna, M. Fortune, C. Fowler, E. Fransson, L. Frassinetti, M. Freisinger, R. Fresa, R. Fridström, D. Frigione, T. Fülöp, M. Furseman, V. Fusco, S. Futatani, D. Gadariya, K. Gal, D. Galassi, K. Galazka, S. Galeani, D. Gallart, R. Galvão, Y. Gao, J. Garcia, M. García-Muñoz, M. Gardener, L. Garzotti, J. Gaspar, R. Gatto, P. Gaudio, D. Gear, T. Gebhart, S. Gee, M. Gelfusa, R. George, S.N. Gerasimov, G. Gervasini, M. Gethins, Z. Ghani, M. Gherendi, F. Ghezzi, J.C. Giacalone, L. Giacomelli, G. Giacometti, C. Gibson, K.J. Gibson, L. Gil, A. Gillgren, D. Gin, E. Giovannozzi, C. Giroud, R. Glen, S. Glöggler, J. Goff, P. Gohil, V. Goloborodko, R. Gomes, B. Gonçalves, M. Goniche, A. Goodyear, S. Gore, G. Gorini, T. Görler, N. Gotts, R. Goulding, E. Gow, B. Graham, J.P. Graves, H. Greuner, B. Grierson, J. Griffiths, S. Griph, D. Grist, W. Gromelski, M. Groth, R. Grove, M. Gruca, D. Guard, N. Gupta, C. Gurl, A. Gusarov, L. Hackett, S. Hacquin, R. Hager, L. Hägg, A. Hakola, M. Halitovs, S. Hall, S.A. Hall, S. Hallworth-Cook, C.J. Ham, D. Hamaguchi, M. Hamed, C. Hamlyn-Harris, K. Hammond, E. Harford, J.R. Harrison, D. Harting, Y. Hatano, D.R. Hatch, T. Haupt, J. Hawes, N.C. Hawkes, J. Hawkins, T. Hayashi, S. Hazael, S. Hazel, P. Heesterman, B. Heidbrink, W. Helou, O. Hemming, S.S. Henderson, R.B. Henriques, D. Hepple, J. Herfindal, G. Hermon, J. Hill, J.C. Hillesheim, K. Hizanidis, A. Hjalmarsson, A. Ho, J. Hobirk, O. Hoenen, C. Hogben, A. Hollingsworth, S. Hollis, E. Hollmann, M. Holzl, B. Homan, M. Hook, D. Hopley, J. Horáček, D. Horsley, N. Horsten, A. Horton, L.D. Horton, L. Horvath, S. Hotchin, R. Howell, Z. Hu, A. Huber, V. Huber, T. Huddleston, G.T.A. Huijsmans, P. Huynh, A. Hynes, M. Iliasova,

D. Imrie, M. Imríšek, J. Ingleby, P. Innocente, K. Insulander Björk, N. Isernia, I. Ivanova-Stanik, E. Ivings,

S. Jablonski, S. Jachmich, T. Jackson, P. Jacquet, H. Järleblad, F. Jaulmes, J.Jenaro Rodriguez, I. Jepu,

E. Joffrin, R. Johnson, T. Johnson, J. Johnston, C. Jones, G. Jones, L. Jones, N. Jones, T. Jones, A. Joyce,

R. Juarez, M. Juvonen, P. Kalnina, T. Kaltiaisenaho, J. Kaniewski, A. Kantor, A. Kappatou, J. Karhunen,

D. Karkinsky, Yu Kashchuk, M. Kaufman, G. Kaveney, Ye.O. Kazakov, V. Kazantzidis, D.L. Keeling, R. Kelly,

M. Kempenaars, C. Kennedy, D. Kennedy, J. Kent, K. Khan, E. Khilkevich, C. Kiefer, J. Kilpelainen, C. Kim, Hyun-Tae Kim, S.H. Kim, D.B. King, R. King, D. Kinna, V.G. Kiptily, A. Kirjasuo, K.K. Kirov, A. Kirschner, T. kiviniemi, G. Kizane, M. Klas, C. Klepper, A. Klix, G. Kneale, M. Knight, P. Knight, R. Knights, S. Knipe, M. Knolker, S. Knott, M. Kocan, F. Köchl, I. Kodeli, Y. Kolesnichenko, Y. Kominis, M. Kong, V. Korovin, B. Kos, D. Kos, H.R. Koslowski, M. Kotschenreuther, M. Koubiti, E. Kowalska-Strzeciwilk, K. Koziol, A. Krasilnikov, V. Krasilnikov, M. Kresina, K. Krieger, N. Krishnan, A. Krivska, U. Kruezi, I. Ksiazek, A.B. Kukushkin, H. Kumpulainen, T. Kurki-Suonio, H. Kurotaki, S. Kwak, O.J. Kwon, L. Laguardia, E. Lagzdina, A. Lahtinen, A. Laing, N. Lam, H.T. Lambertz, B. Lane, C. Lane, E.Lascas Neto, E. Laszyńska, K.D. Lawson, A. Lazaros, E. Lazzaro, G. Learoyd, Chanyoung Lee, S.E. Lee, S. Leerink, T. Leeson, X. Lefebvre, H.J. Leggate, J. Lehmann, M. Lehnen, D. Leichtle, F. Leipold, I. Lengar, M. Lennholm, E. Leon Gutierrez, B. Lepiavko, J. Leppänen, E. Lerche, A. Lescinskis, J. Lewis, W. Leysen, L. Li, Y. Li, J. Likonen, Ch. Linsmeier, B. Lipschultz, X. Litaudon, E. Litherland-Smith, F. Liu, T. Loarer, A. Loarte, R. Lobel, B. Lomanowski, P.J. Lomas, J.M. López, R. Lorenzini, S. Loreti, U. Losada, V.P. Loschiavo, M. Loughlin, Z. Louka, J. Lovell, T. Lowe, C. Lowry, S. Lubbad, T. Luce, R. Lucock, A. Lukin, C. Luna, E.de la Luna, M. Lungaroni, C.P. Lungu, T. Lunt, V. Lutsenko, B. Lyons, A. Lyssoivan, M. Machielsen, E. Macusova, R. Mäenpää, C.F. Maggi, R. Maggiora, M. Magness, S. Mahesan, H. Maier, R. Maingi, K. Malinowski, P. Manas, P. Mantica, M.J. Mantsinen, J. Manyer, A. Manzanares, Ph. Maquet, G. Marceca, N. Marcenko, C. Marchetto, O. Marchuk,

A. Mariani, G. Mariano, M. Marin, M. Marinelli, T. Markovič, D. Marocco, L. Marot, S. Marsden, J. Marsh, R. Marshall, L. Martellucci, A. Martin, A.J. Martin, R. Martone, S. Maruyama, M. Maslov, S. Masuzaki, S. Matejcik, M. Mattei, G.F. Matthews, D. Matveev, E. Matveeva, A. Mauriya, F. Maviglia, M. Mayer, M.-L. Mayoral, S. Mazzi, C. Mazzotta, R. McAdams, P.J. McCarthy, K.G. McClements, J. McClenaghan, P. McCullen, D.C. McDonald, D. McGuckin, D. McHugh, G. McIntyre, R. McKean, J. McKehon, B. McMillan,

L. McNamee, A. McShee, A. Meakins, S. Medley, C.J. Meekes, K. Meghani, A.G. Meigs, G. Meisl, S. Meitner,

S. Menmuir, K. Mergia, S. Merriman, Ph. Mertens, S. Meshchaninov, A. Messiaen, R. Michling, P. Middleton,

D. Middleton-Gear, J. Mietelski, D. Milanesio, E. Milani, F. Militello, A.Militello Asp, J. Milnes, A. Milocco,

G. Miloshevsky, C. Minghao, S. Minucci, I. Miron, M. Miyamoto, J. Mlynář, V. Moiseenko, P. Monaghan,

I. Monakhov, T. Moody, S. Moon, R. Mooney, S. Moradi, J. Morales, R.B. Morales, S. Mordijck, L. Moreira,

L. Morgan, F. Moro, J. Morris, K.-M. Morrison, L. Msero, D. Moulton, T. Mrowetz, T. Mundy, M. Muraglia,

A. Murari, A. Muraro, N. Muthusonai, B. N'Konga, Yong-Su Na, F. Nabais, M. Naden, J. Naish, R. Naish,

F. Napoli, E. Nardon, V. Naulin, M.F.F. Nave, I. Nedzelskiy, G. Nemtsev, V. Nesenevich, I. Nestoras, R. Neu,

V.S. Neverov, S. Ng, M. Nicassio, A.H. Nielsen, D. Nina, D. Nishijima, C. Noble, C.R. Nobs, M. Nocente,

## Acknowledgments

J. Garcia would like to thank Gerardo Giruzzi for fruitful discussions.

This work has been carried out within the framework of the EUROfusion Consortium, funded by the European Union via the Euratom Research and Training Programme (Grant Agreement No 101052200 - EUROfusion) and from the EPSRC [grant number EP/W006839/1]. Views and opinions expressed are however those of the author(s) only and do not necessarily reflect those of the European Union or the European Commission. Neither the European Union nor the European Commission can be held responsible for them.

This work was partly supported by grants FIS2017-85252-R and PID2021-127727OB-I00, funded by the Spanish Ministry of Science and Innovation.

An award of computer time was provided by the INCITE program and ALCC program. This research used resources from the Oak Ridge Leadership Computing Facility, which is an Office of Science User Facility supported under Contract DE-AC05-00OR22725. Computing resources were also provided by the National Energy Research Scientific Computing Center, which is an Office of Science User Facility supported under Contract DEAC0205CH11231.

## Author contributions

The reported experiments were devised and jointly led by Y.K., J.O., S.S., J.G. and M.B., with the key coordination of E. de la L., C.F.M and J.M. The TRANSP simulations were performed by Z.S. and M.P. Gyrokinetic simulations and subsequent analyses were performed by E.B, J.C. and S.M. FAR3D simulations were performed by J.V. with the assistance of D.Z. Reflectometer analyses were performed by M.D. and J.R.R. MHD analyses were performed by R.C. and M.D. Pedestal analyses were performed by E.de la L. and E.S. Alpha particle losses were investigated by P.J.B. The manuscript was written by J.G. and E. de la L. with the feedback by all the authors.

## Competing interests

The authors declare no competing interests.

## Additional Information

Correspondence and requests for materials should be addressed to J. Garcia.